\documentclass[a4paper,11pt]{article}
\usepackage{pos}
\usepackage{tikz}
\usepackage{calc}
\usetikzlibrary{shapes,arrows}
\usetikzlibrary{fit,calc}
\usetikzlibrary{positioning}

\newcommand{\pp}{\ensuremath{\text{p\kern-0.05em p}}}

\newcommand{\PbPb}{\ensuremath{\mbox{Pb--Pb}}}

\newcommand{\sqrts}{\ensuremath{\sqrt{s_{\text{NN}}}}}
\newcommand{\sqrtsNoNN}{\ensuremath{\sqrt{s}}}

\newcommand{\GeVc}{\ensuremath{\text{GeV}\kern-0.05em/\kern-0.02em c}}

\newcommand{\pT}{\ensuremath{p_{\text{T}}}}

\newcommand{\pTJet}{\ensuremath{p_{\text{T,jet}}}}

\newcommand{\pTJetCh}{\ensuremath{p_{\text{T,jet}}^{\text{ch}}}}

\newcommand{\zcut}{\ensuremath{z_{\text{cut}}}}

\newcommand{\kT}{\ensuremath{k_{\text{T}}}}
\newcommand{\kTg}{\ensuremath{k_{\text{T,g}}}}

\newcommand{\moliere}{Moli\`ere}

\title{Exploring medium properties with hard transverse momentum splittings using groomed jet substructure measurements in Pb--Pb collisions with ALICE}
\ShortTitle{Exploring medium properties via jet substructure with ALICE}

\manuallySeparateAuthors{}
\author*{Raymond Ehlers}
\author{on behalf of the ALICE Collaboration}

\affiliation{Lawrence Berkeley National Laboratory/UC Berkeley}

\emailAdd{raymond.ehlers@cern.ch}

\abstract{Jet substructure observables provide unique probes of the properties of the Quark-Gluon Plasma (QGP).
In these proceedings we report new measurements of groomed jet substructure in central Pb--Pb collisions at $\sqrt{s_\mathrm{NN}}=5.02$ TeV.
We present the first application of dynamical grooming in heavy-ion collisions to search for excess $k_{\mathrm{T,g}}$ emissions, which is a signature of large-angle scattering of jets off quasi-particles in the QGP.
We present additional measurements employing both the soft drop and dynamical grooming algorithms in 0--10\% central Pb--Pb, 30--50\% semicentral Pb--Pb, and proton--proton (pp) collisions at $\sqrt{s_{\mathrm{NN}}}=5.02$ TeV.
Results from the various grooming methods and parameters are compared.
Comparisons to model calculations are also presented.
The techniques developed for this measurement are more broadly applicable to jet substructure, which we likewise discussed.}

\FullConference{HardProbes2023\\
 26-31 March 2023\\
 Aschaffenburg, Germany\\}

\begin{document}
\maketitle

\section{Introduction}

Jets are collimated sprays of particles produced by high-momentum transfer processes in high-energy nuclear collisions.
These objects are multi-scale probes of the collision dynamics, providing insight into perturbative Quantum Chromodynamics (QCD) as well as the QCD medium formed in heavy-ion collisions known as the quark-gluon plasma (QGP).
As jets propagate through the QGP, their interactions with the medium encode properties of the medium itself via modification of jet properties and structure, called jet quenching.
Measurements of jet substructure modification, and their comparison with theoretical models, may resolve different medium length scales, in particular those of emergent quasi-particle structure~\cite{DEramo:2013aa,DEramo:2018eoy}, or the critical angle for resolving color charges in the medium.
In these proceedings, ALICE reports a new jet substructure measurement, searching for excess high $\kT{}$ emissions in $\PbPb{}$ collisions as a signature of point-like (\moliere{}) scattering in the medium, in analogy to Rutherford scattering~\cite{DEramo:2013aa,DEramo:2018eoy}.
This approach is complementary to the search for such scattering centers using jet deflection, as explored by ALICE and STAR~\cite{Adam:2015doa,STAR:2017hhs}.

The ALICE apparatus~\cite{Abelev:2014ffa} is well suited to these substructure measurements because of its ability to measure low $\pT{}$ jets and small splitting angles with high efficiency.
Low $\pT{}$ jets are of particular interest since they should be more sensitive to jet deflection due to scattering centers.
The analysis was performed using data collected from $\pp{}$ and $\PbPb{}$ collision at $\sqrt{s_{\text{NN}}}$ = 5.02 TeV during the 2017 and 2018 LHC runs.
Charged-particle jets are found using charged tracks as input to the anti-$\kT{}$ jet finding algorithm implemented in FastJet~\cite{Cacciari:2011ma} with jet resolution parameter $R$ = 0.2.
Jet candidates are required to be contained fully within the fiducial acceptance, with background contributions in \(\PbPb{}\) collisions subtracted via event-wise constituent subtraction
\cite{Berta:2019aa}.

After jet finding and selection, jet constituents are then reclustered with the Cambridge/Aachen (C/A) algorithm.
The reclustered splitting tree is then unwound, with each identified splitting evaluated according to a grooming algorithm, which aims to remove soft, wide-angle emissions to isolate the hardest splittings.
We iteratively follow the harder subjet through the splitting tree, evaluating the grooming condition for each splitting.
Grooming algorithms make selections within the Lund Jet Plane, emphasizing particular regions of phase space.
In these proceedings, observables are measured utilizing either 1) the dynamical grooming (DyG) algorithm with dynamic hardness measure \(a\), which can be used to focus on, e.g.,~the largest \(\kT{}\) (\(a = 1.0\))~\cite{Mehtar-Tani:2019rrk}, or 2) the soft drop (SD) grooming algorithm with a minimum shared momentum fraction requirement, \(\zcut{}\)~\cite{Larkoski:2014wba}.
The iterative splitting with the largest value of the hardness measure (in DyG) or the first iterative splitting which passes the condition (in SD) is selected, and the $\kT{}$ value is calculated and recorded as $\kTg{}$.
All observables are corrected for background fluctuations and detector effects via two-dimensional unfolding ($\pTJet{}$, $\kTg{}$) using Bayesian iterative unfolding, as implemented in RooUnfold~\cite{Adye:1349242}.

\section{Characterizing grooming methods}

\begin{figure}[t]
    \centering
    \includegraphics[width=0.48\textwidth]{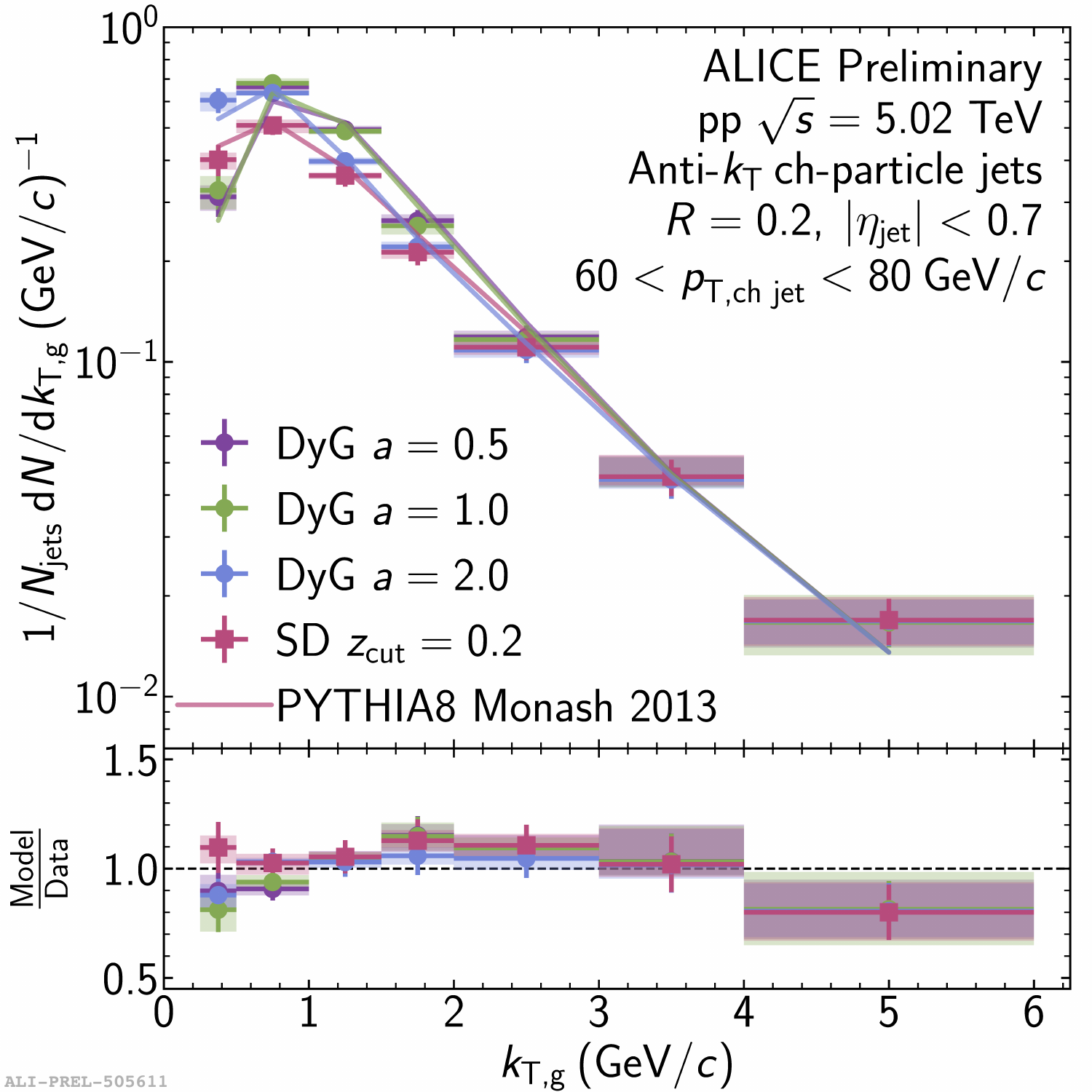}
    \includegraphics[width=0.48\textwidth]{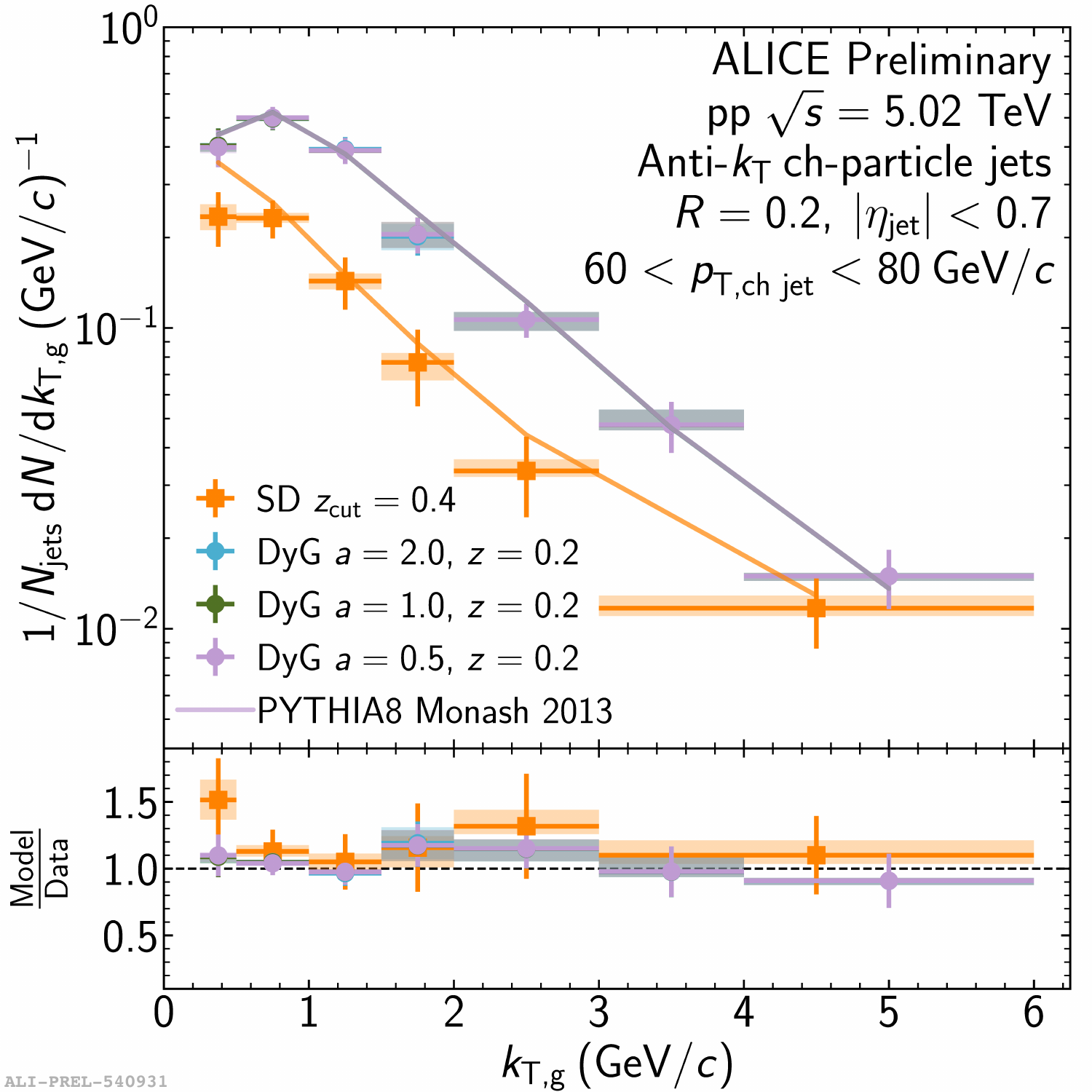}
    \caption{$\kTg{}$ spectra measured using a variety of grooming methods for charged-particle jets with $R = 0.2$ in $60 < \pTJetCh{} < 80$ \GeVc{} in $\pp{}$ collisions at $\sqrtsNoNN{}$ = 5.02 TeV. The spectra are compared with PYTHIA 8 calculations.}
    \label{fig:ppGroomingMethods}
\end{figure}

Due to the contribution of underlying event background on the reconstructed splitting tree, the optimal grooming algorithm to identify the hardest $\kT{}$ splittings is not known a priori.
To investigate the impact of grooming, a collection of algorithms was investigated: 1) SD with $\zcut{}$ = 0.2 and 0.4, 2) DyG with $a$ = 0.5, 1.0, 2.0, and 3) DyG with $a$ = 0.5, 1.0, 2.0 for all splittings with $z > 0.2$.
Dynamical grooming with the $z$ requirement provides a link to the soft drop algorithm, as well as aiding in expanding the measured $\kTg{}$ range, as described in further detail below.
The results obtained with the different grooming algorithms are compared first in $\pp{}$ collisions via the $\kTg{}$ spectra measured for $R$ = 0.2 jets with $60 < \pTJetCh{} < 80$ \GeVc{} reconstructed in $\pp{}$ collisions at $\sqrtsNoNN{}$ = 5.02 TeV.
The spectra are shown in Fig.~\ref{fig:ppGroomingMethods}.
At high $\kT{}$, all grooming methods converge on a consistent set of splittings except for soft drop $\zcut{}$ = 0.4, where the overall yield is reduced due to the limited phase space available given the $z$ requirement.
However, at low to intermediate $\kT{}$, differences between the methods emerge, particularly for the dynamical grooming methods without a $z$ requirement.
The spectra are generally compatible with expectations from PYTHIA 8~\cite{Sjostrand:2015aa} using the Monash 2013 tune~\cite{Skands:2014pea}, although there are some hints of tension in the shape.

Moving to $\PbPb{}$ collisions, stable unfolding requires sufficiently high probability of matching leading and subleading subjet prongs when constructing the response matrix~\cite{ALargeIonColliderExperiment:2021mqf}.
For dynamical grooming, this is difficult to achieve because of background contributions at low $\kT{}$ for jets that only undergo relatively soft splittings, leading to reduced subleading subjet purity and off-diagonal mismatched splittings.
This can be mitigated with a minimum measured $\kT{}$ requirement, trading off reduced kinematic efficiency and thus $\kT{}$ dynamic range for higher purity.
This issue can alternatively be addressed by a minimum $z$ requirement, as in soft drop or dynamical grooming with a $z$ requirement (labeled as case 3 above).
The techniques described here can be applied to other jet substructure measurements, enabling the application of dynamical grooming or other variations of grooming methods to other observables and regions of phase space which were not previously within reach.

\begin{figure}[t]
    \centering
    \includegraphics[width=0.48\textwidth]{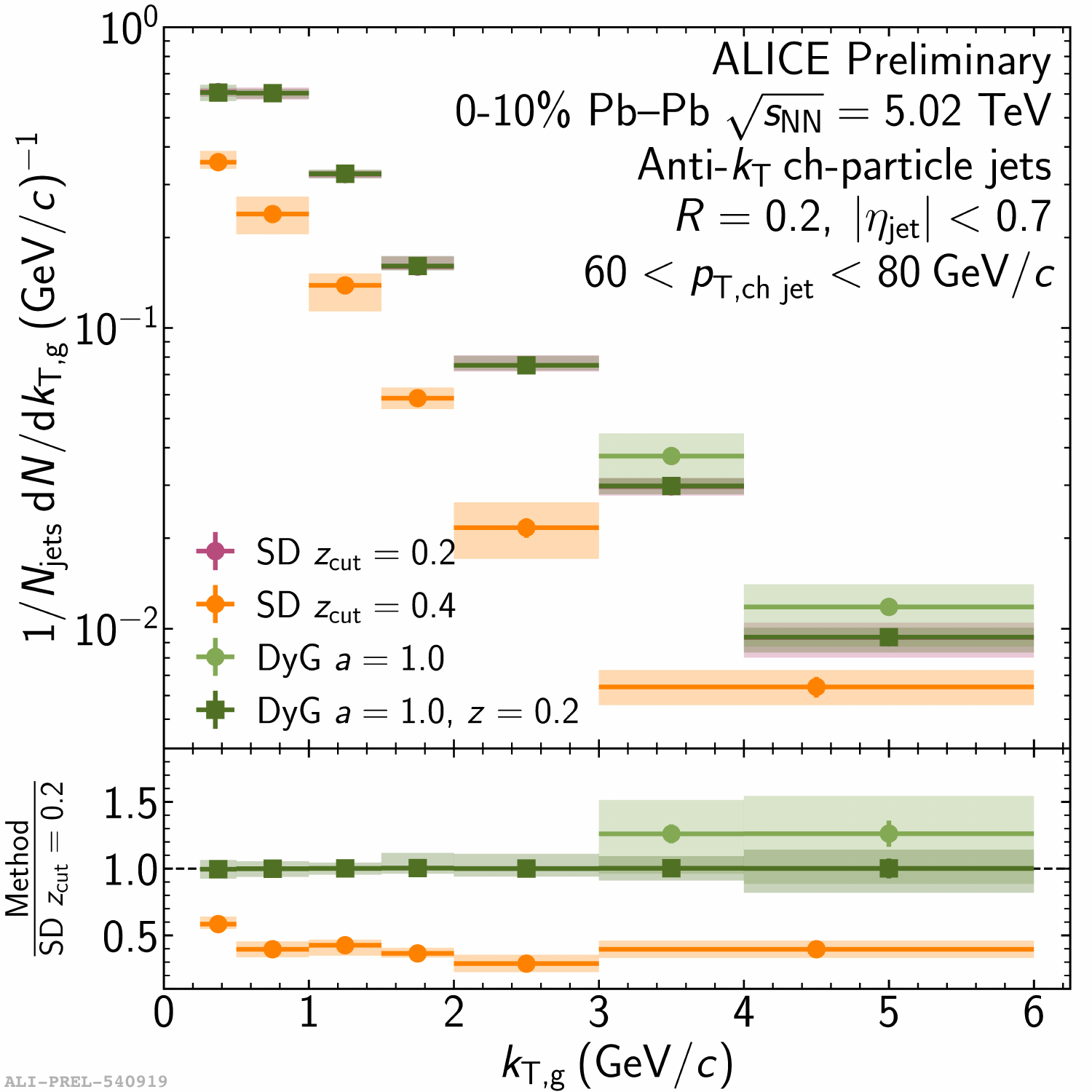}
    \includegraphics[width=0.48\textwidth]{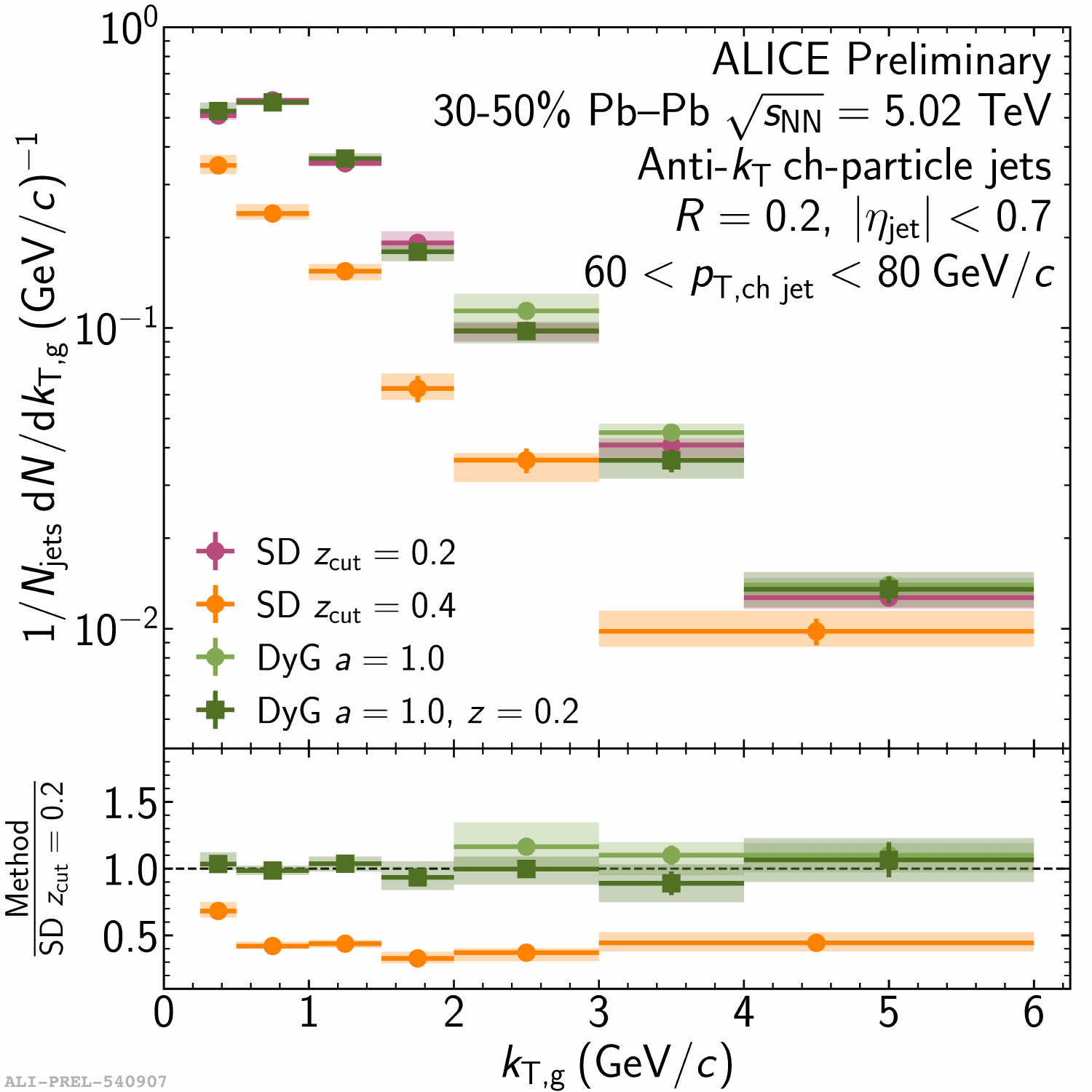}
    \caption{$\kTg{}$ spectra measured using a selection of grooming methods for charged-particle jets with $R = 0.2$ in $60 < \pTJetCh{} < 80$ \GeVc{} in 0--10\% central (left) and 30--50\% semi-central (right) $\PbPb{}$ collisions.}
    \label{fig:PbPbGroomingMethodsComparison}
\end{figure}

Using this approach, ALICE performed the first measurements using dynamical grooming in heavy ion collisions, which are shown in Fig.~\ref{fig:PbPbGroomingMethodsComparison} along with a selection of other grooming methods.
The $R = 0.2$ jets with transverse momentum $60 < \pTJetCh{} < 80$ \GeVc{} are reconstructed in 0--10\% (left) and 30--50\% (right) $\PbPb{}$ collisions at $\sqrts{}$ = 5.02 TeV.
Similar trends in the grooming methods are seen for both centrality ranges, matching the trends seen in $\pp{}$ collisions.
The minimum measured $\kT{}$ requirement for standard dynamical grooming restricts the range which can be reported, with the smaller background in 30--50\% enabling an additional intermediate $\kTg{}$ bin (2-3 \GeVc{}).
The yield of DyG $a$ = 1.0, DyG $a$ = 1.0 with $z > 0.2$, and SD $\zcut{}$ = 0.2 are consistent at high $\kT{}$ within systematic and statistical uncertainties, while SD $\zcut{}$ = 0.4 is again suppressed due to phase space selections.
Varying the dynamical grooming hardness measure $a$ is consistent within uncertainties in the reported range (not shown).
The consistency of so many methods suggests that all of the methods are selecting the same set of splittings at high $\kTg{}$.
Further, the consistency between soft drop $\zcut{}$ = 0.2 (which selects the first splitting that passes the grooming condition) and dynamical grooming with a $z$ requirement (which considers all iterative splittings passing the $z$ requirement) suggests that it is rare for a harder splitting to occur further into the splitting tree for jets measured within the considered kinematic range.

\section{Search for quasi-particle scattering}

\begin{figure}[t]
    \centering
    \includegraphics[width=0.32\textwidth]{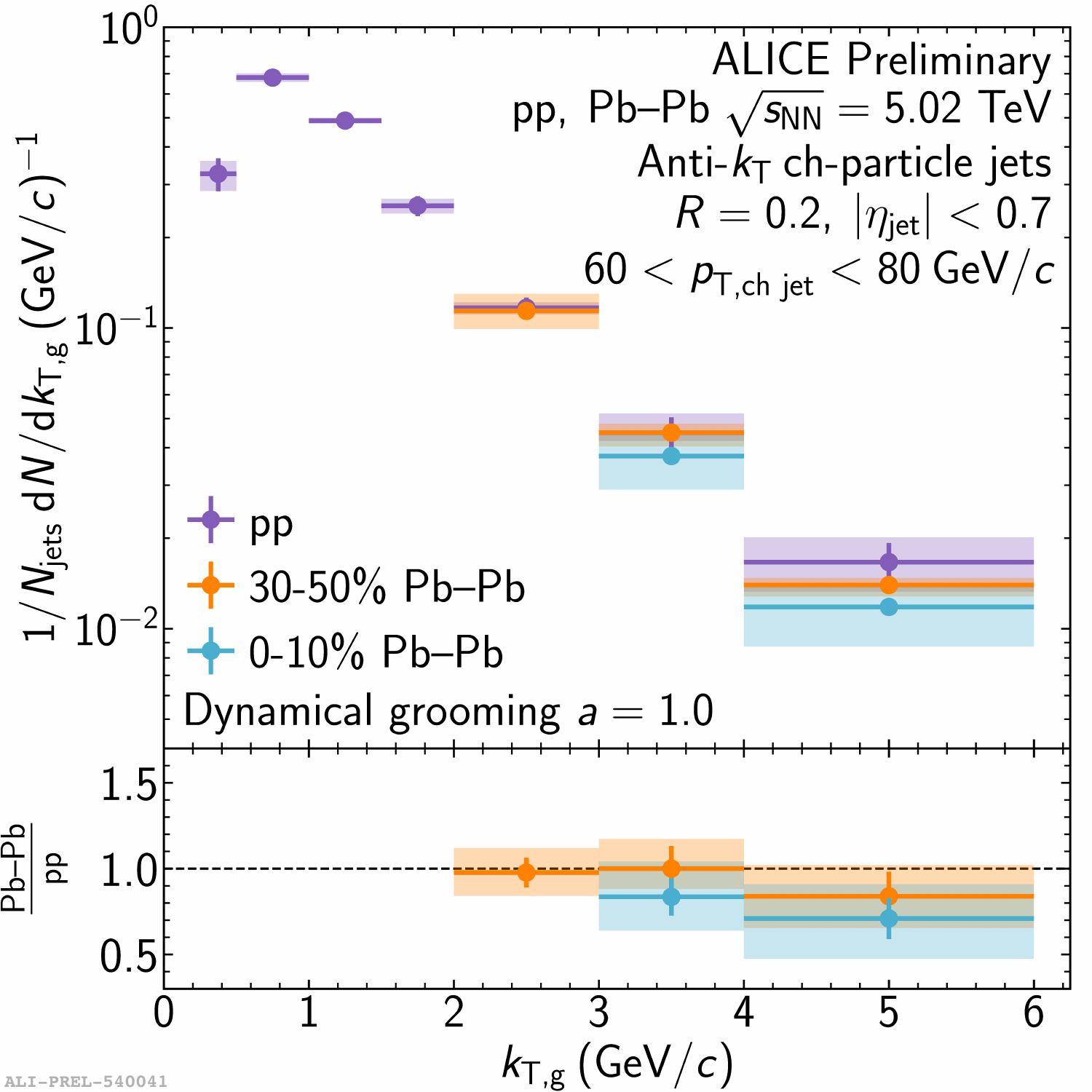}
    \includegraphics[width=0.32\textwidth]{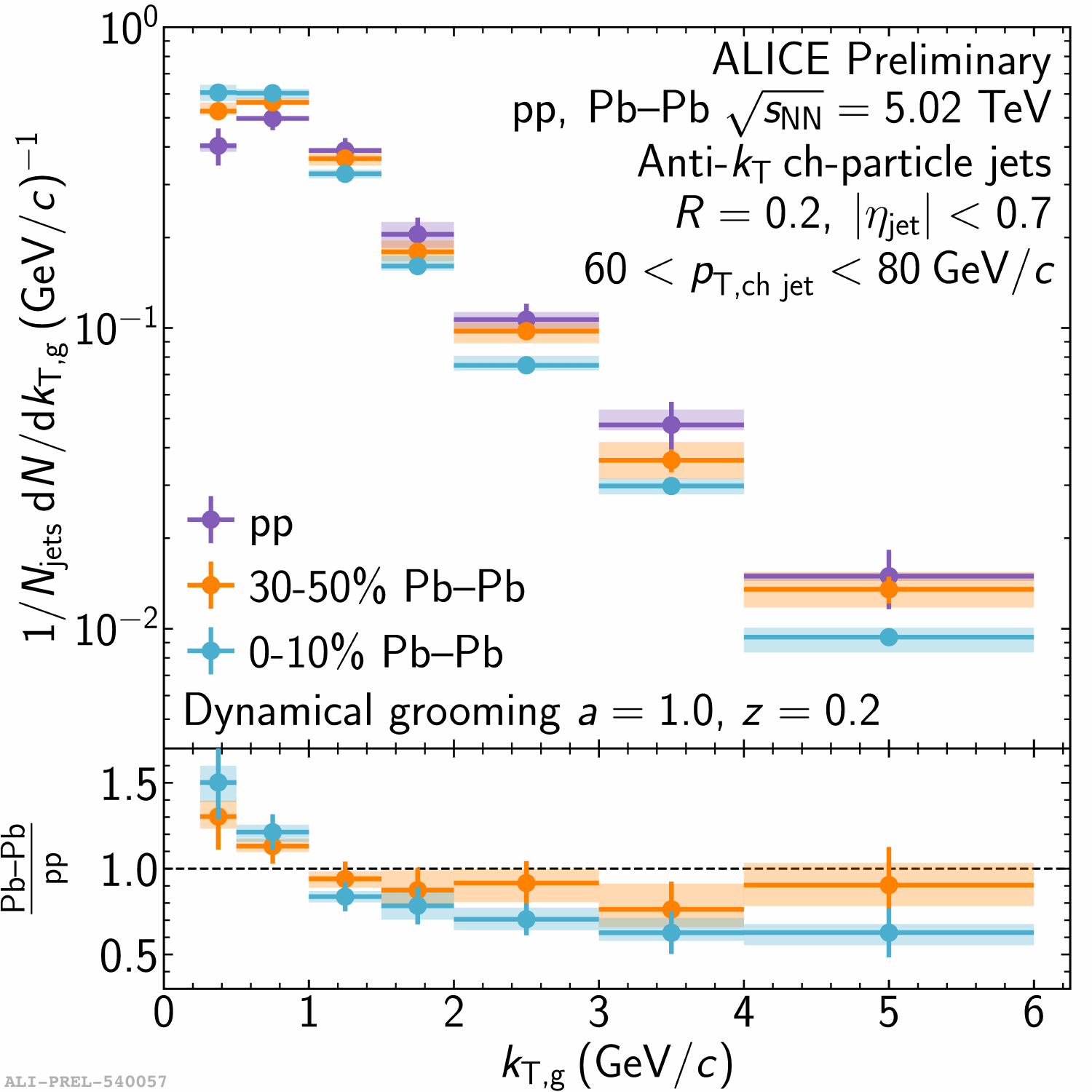}
    \includegraphics[width=0.32\textwidth]{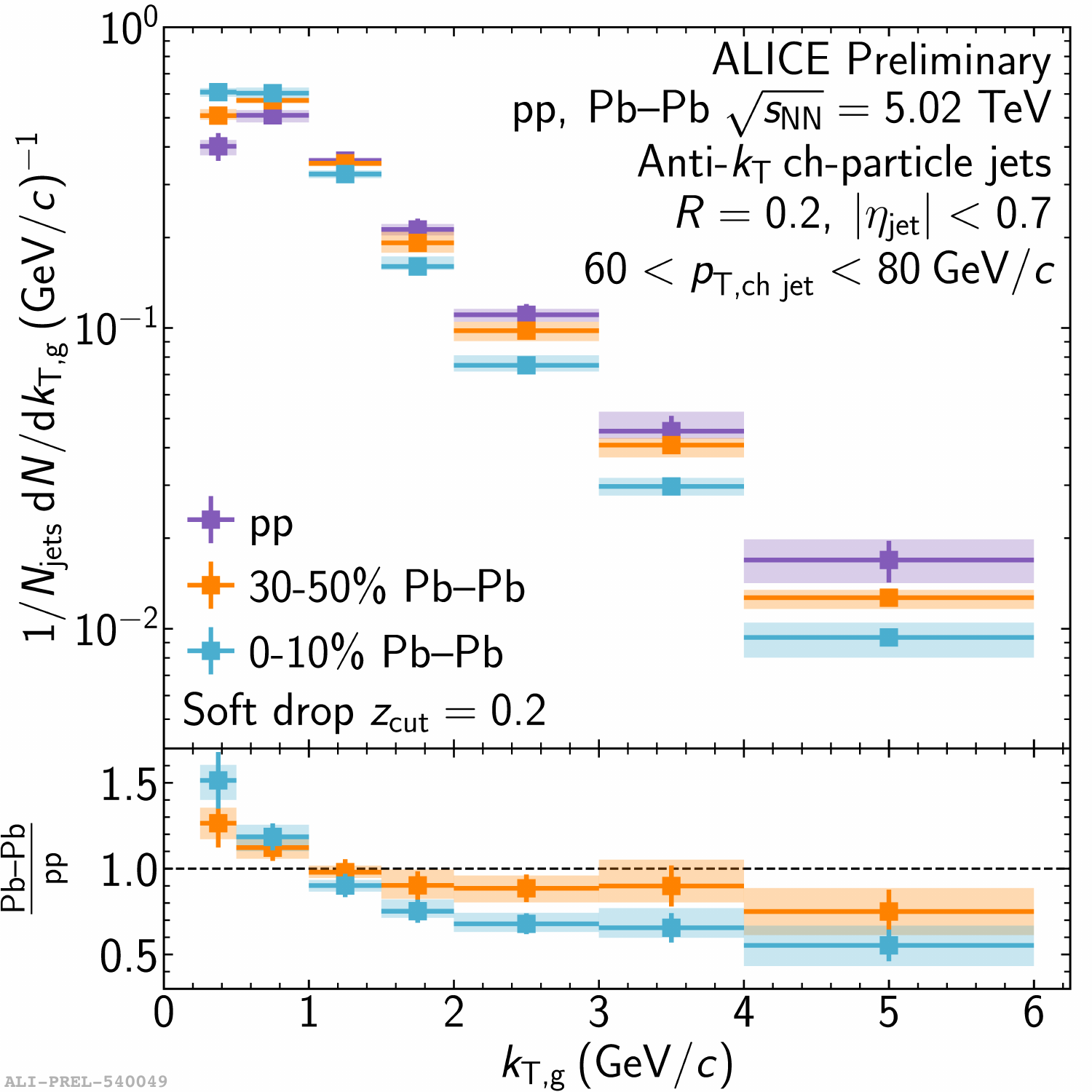}
    \caption{The $\kTg{}$ spectra (upper panels) measured for charged-particle jets with $R = 0.2$ in $60 < \pTJetCh{} < 80$ \GeVc{} in 0--10\% central, 30--50\% semi-central $\PbPb{}$ collisions, and $\pp{}$ collisions and their ratio to $\pp{}$ collisions (lower panels) using dynamical grooming $a$ = 1.0 (left), dynamical grooming $a$ = 1.0 with $z > 0.2$ (middle), and soft drop $\zcut{}$ = 0.2 (right).}
    \label{fig:PbPbMoliereSearch}
\end{figure}

With guidance from the grooming method characterization studies, we then turn our attention to possible signals of point-like (\moliere{}) scattering by comparing the $\kTg{}$ yield in $\PbPb{}$ to that in $\pp{}$ collisions.
The results of this comparison are shown in Fig.~\ref{fig:PbPbMoliereSearch} for DyG $a$ = 1.0 (left), DyG $a$ = 1.0 with $z > 0.2$ (middle), and SD $\zcut{}$ = 0.2 (right).
For all three methods, the $\kTg{}$ spectra is modified in $\PbPb{}$ collisions, although there is no evidence of enhancement at high $\kTg{}$.
For DyG $a$ = 1.0, there is a hint of suppression at high $\kTg{}$, although the $\PbPb{}$ to $\pp{}$ ratio is consistent with unity within uncertainties.
The cases of DyG $a$ with $z > 0.2$ and SD $\zcut{}$ = 0.2 show trends that are consistent with each other, illustrating enhancement at low $\kTg{}$ and suppression at high $\kTg{}$.
The size of the modification is larger in 0--10\% compared to 30--50\% collisions, which is consistent with jet energy loss expectations.
Indeed, this trend is consistent with those previously measured in angular-dependent substructure observables~\cite{ALargeIonColliderExperiment:2021mqf}.

\begin{figure}[ht]
    \centering
    \includegraphics[width=0.48\textwidth]{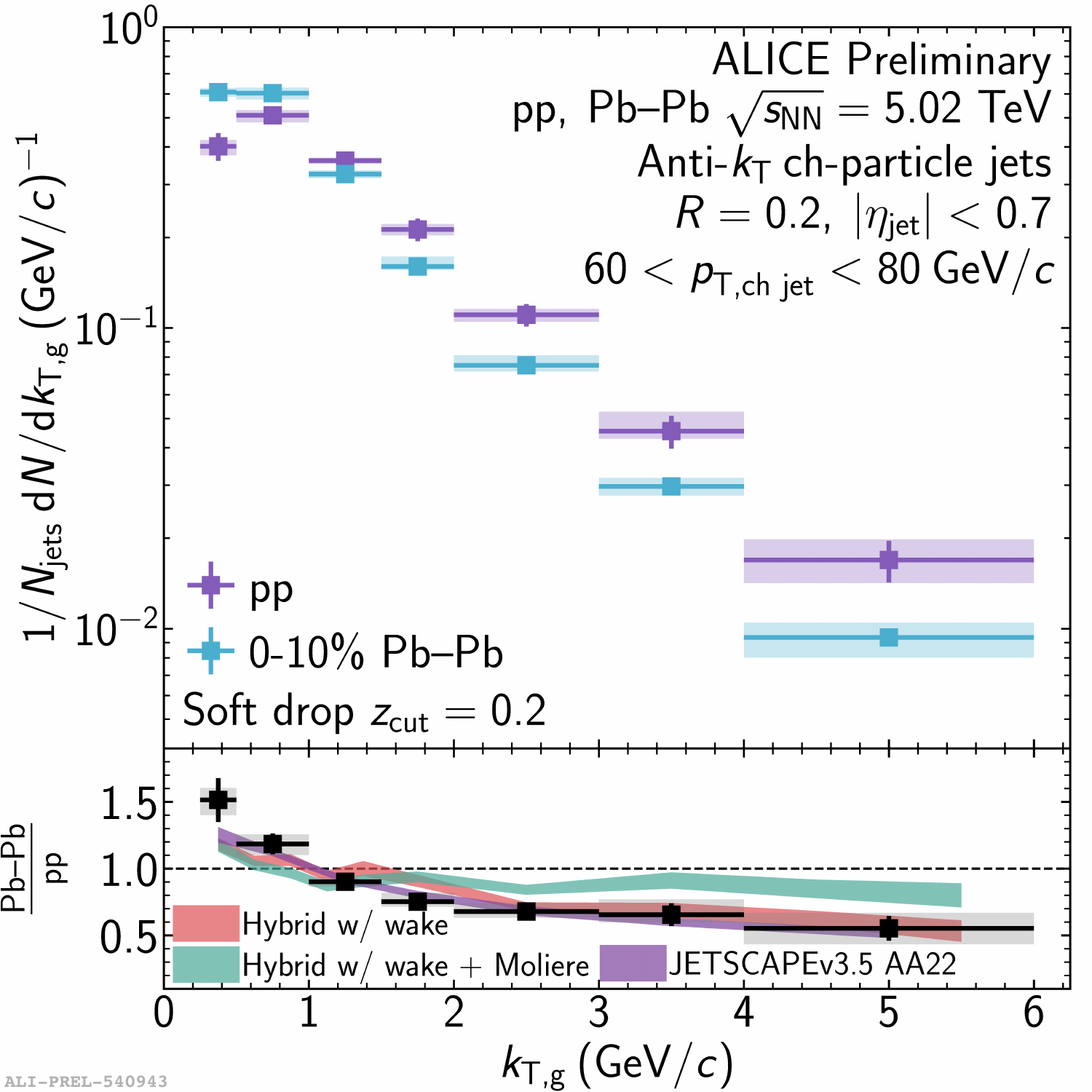}
    \includegraphics[width=0.48\textwidth]{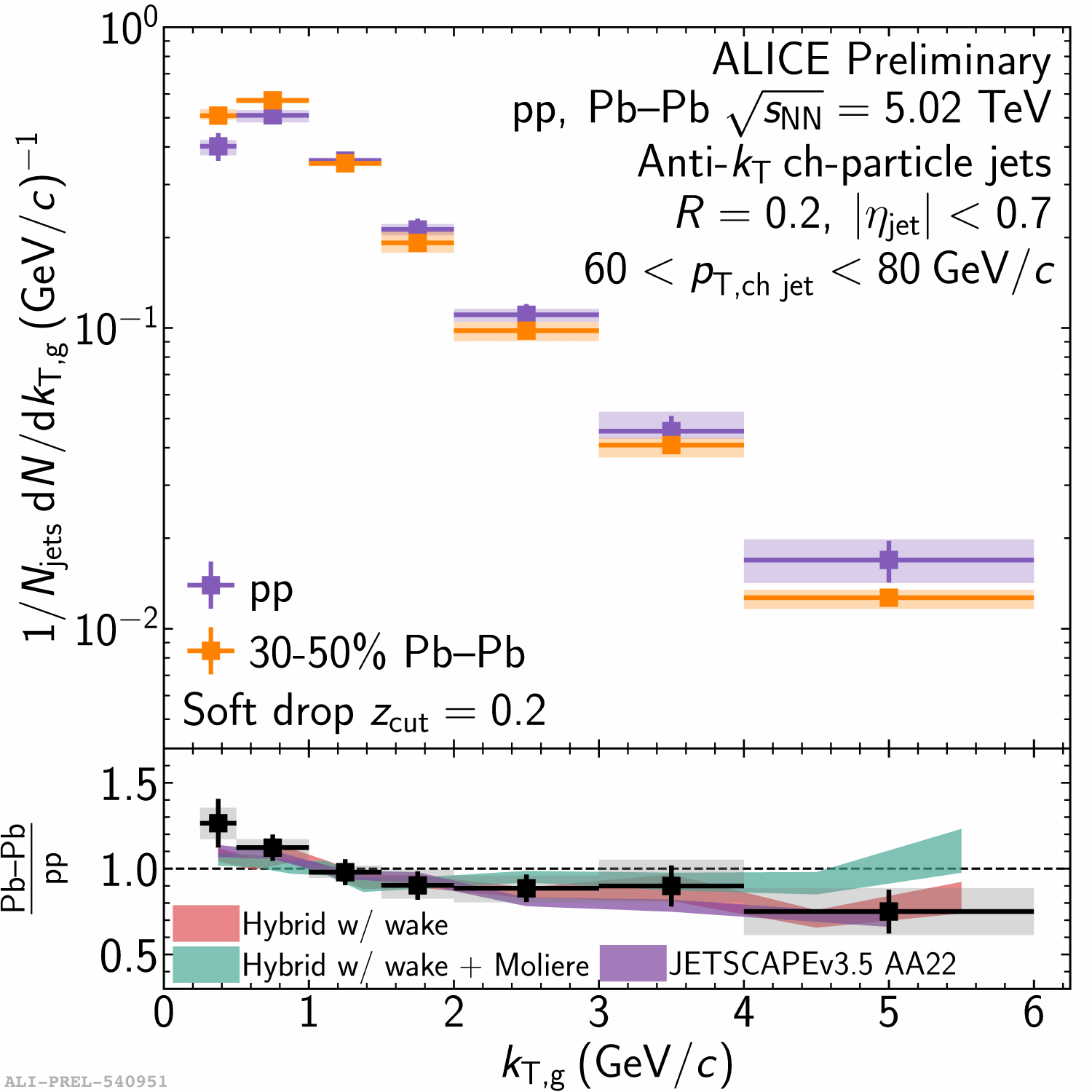}
    \caption{The $\kTg{}$ spectra measured for charged-particle jets with $R = 0.2$ in $60 < \pTJetCh{} < 80$ \GeVc{} in 0--10\% central (left), 30--50\% semi-central $\PbPb{}$ (right) collisions and $\pp{}$ collisions (upper) and their modification relative to $\pp{}$ collisions using soft drop $\zcut{}$ = 0.2 are compared to model calculations from JETSCAPE and the Hybrid model.}
    \label{fig:PbPbModelComparisons}
\end{figure}

Though there is no clear enhancement at high $\kTg{}$ in the spectra, the impact of point-like scattering may still be present in the data, convolved with the suppression due to the energy loss.
To gain further insight into the measurement, the spectra are compared to model calculations, including: 
JETSCAPE v3.5 (AA22 tune)~\cite{JETSCAPE:2022jer,JETSCAPE:2023hqn}, which utilizes a multi-stage approach with MATTER in the high-virtuality phase and LBT in the low-virtuality phase;
and the Hybrid model~\cite{DEramo:2018eoy}, which implements an AdS-CFT inspired approach to energy loss and includes the ability to isolate the \moliere{} scattering contribution.
These models are compared to the ratio of the $\kTg{}$ spectra measured in $\PbPb{}$ and $\pp{}$ collisions in Fig.~\ref{fig:PbPbModelComparisons}.
The JETSCAPE calculations include \moliere{} scattering, and describe the ratio well in both 0--10\% and 30--50\% $\PbPb{}$ collisions.
While the Hybrid model calculations show that \moliere{} scattering is expected to have an impact in this kinematic region, the 0--10\% ratio appear to show a preference for the calculations without the \moliere{} component.
However, this conclusion is also dependent on how well the model describes the underlying $\pp{}$ and $\PbPb{}$ spectra, which remains to be studied.

\section{Outlook}

ALICE has performed a search for evidence of the quasi-particle nature of the quark-gluon plasma via measurements of the relative transverse momentum of jet splittings, $\kTg{}$.
Utilizing a collection of grooming methods - including dynamical grooming for the first time in heavy-ion collisions - the spectra were measured for $R = 0.2$ charged-particle jets in $\pp{}$ and $\PbPb{}$ collisions at $\sqrts{}$ = 5.02 TeV.
The performance of soft drop and dynamical grooming with and without a minimum $z$ requirement was characterized in all collision systems, demonstrating that a consistent set of high $\kTg{}$ splittings are selected by most algorithms.
Furthermore, the $z$ requirement was shown to dominate over details of the method, suggesting that there are few hard splittings to be found in the jet splitting tree beyond the first hard splitting.
The $\kTg{}$ spectra are modified in $\PbPb{}$ collisions relative to $\pp{}$, similarly to other angular-dependent substructure variables, with the magnitude of the modification depending on the grooming method and centrality.
There is no clear evidence of \moliere{} scattering at high $\kTg{}$ in the measured kinematic range, although a possible enhancement relative to the overall suppression driven by jet energy loss driven is not yet excluded.

\scriptsize
\bibliographystyle{style/JHEP}
\bibliography{rehlers_HP2023.bib}

\end{document}